\documentclass[prl,twocolumn,aps,floatfix,longbibliography,nofootinbib]{revtex4-2}

\makeatletter 
\newcommand\seminormal{\@setfontsize\seminormal{10.46}{12.77}}
\makeatother 
\usepackage[utf8]{inputenc}
\usepackage[T1]{fontenc}
\usepackage{pslatex, graphicx, dcolumn, bm, amssymb, amsmath, mathtools, physics, soul, comment, stackengine}
\usepackage[dvipsnames]{xcolor}

\usepackage[colorlinks=true,
            linkcolor=blue,       
            citecolor=blue,       
            urlcolor=blue,        
            filecolor=blue,
            breaklinks=true]{hyperref}

\newcommand{\be}{\begin{equation}}
\newcommand{\ee}{\end{equation}}

\setstcolor{red}
\stackMath

\begin{document}
\raggedbottom

\title{Noise-robust navigation from an adaptive run-and-tumble policy}

\author{Aniruddha Datta}
\affiliation{School of Physics, Georgia Institute of Technology, Atlanta, GA 30332, USA}
\author{Shiladitya Banerjee}
\email{Correspondence: sbanerjee347@gatech.edu}
\affiliation{School of Physics, Georgia Institute of Technology, Atlanta, GA 30332, USA}

\begin{abstract}
How do organisms navigate when the signals guiding them are noisy? Variance adaptation, the rescaling of sensitivity to noise, is common in sensory systems, but its role in navigation is unexplored. We introduce a minimal active Brownian particle whose run-and-tumble policy follows from an optimality principle. Variance adaptation emerges as part of this policy. Adaptation keeps chemotactic drift finite as noise grows, while a non-adaptive particle's collapses exponentially. Adaptation also carries a cost, degrading performance in quiet environments and requiring a tuned adaptation sensitivity.
\end{abstract}

\maketitle 
Motile organisms navigate their environment to find nutrients and avoid harm, a task complicated by the noise pervading natural signals \cite{adlerChemotaxisBacteriaMotile1966, mesibovChemotaxisAminoAcids1972}. Bacteria such as {\it E. coli} solve this by comparing signal levels over time \cite{bergPhysicsChemoreception1977, segallTemporalComparisonsBacterial1986, celaniBacterialStrategiesChemotaxis2010}, sensing the logarithm of concentration~\cite{kalininLogarithmicSensingEscherichia2009} and modulating their motion accordingly~\cite{tuModelingChemotacticResponse2008}. Receptor methylation lets them adapt this comparison to the mean background concentration, keeping their response sensitive across a several-fold range of concentrations~\cite{bergPhysicsChemoreception1977, segallTemporalComparisonsBacterial1986, barkaiRobustnessSimpleBiochemical1997, sourjikReceptorSensitivityBacterial2002}. Even the accuracy of this sensing is set by noise in the pathway itself~\cite{mattinglyColiChemosensingAccuracy2026}, making robustness to noise, not just detection of a gradient, central to chemotaxis.

Just as adaptation to the mean extends the range over which a sensor operates reliably, adaptation to the variance of a signal has been shown to maximize the information a system transmits about it~\cite{brennerAdaptiveRescalingMaximizes2000, fairhallEfficiencyAmbiguityAdaptive2001}.  This has been demonstrated directly in motion-sensitive neurons of the fly visual system~\cite{brennerAdaptiveRescalingMaximizes2000}, in auditory and visual processing \cite{nagelTemporalProcessingAdaptation2006,kimTemporalContrastAdaptation2001}. Notably, {\it Drosophila} larvae adapt their turning decisions to the variance of sensory input on a timescale consistent with optimal estimation of environmental variability~\cite{gepnerVarianceAdaptationNavigational2018}, directly linking variance adaptation to navigation.

Although variance adaptation has been studied at the single neuron level as a principle of efficient sensory encoding~\cite{brennerAdaptiveRescalingMaximizes2000,fairhallEfficiencyAmbiguityAdaptive2001}, no physical model has examined its functional payoffs for navigation in simple, moving agents. Whether and how such adaptation benefits an agent whose own actions change the signal it senses remains unexplored. Existing models of chemotactic active particles couple velocity directly to the signal gradient~\cite{liebchenSyntheticChemotaxisCollective2018}, leaving any adaptive, noise-dependent control unaddressed.
Here we introduce a minimal active agent whose run-and-tumble policy, derived from an optimality principle, tracks the mean and amplitude of a noisy signal and adapts its motion to the sensed noise. This variance adaptation yields larger chemotactic drift than a non-adaptive agent. Furthermore, the degree to which the agent is sensitive to environmental noise affects its success in chemotaxis in spatially varying noise. Finally, we show that the adaptive run-and-tumble policy lets populations locate hidden sources faster across noisy regions than non-adaptive agents.

\textit{The model. \textemdash} We model the chemotactic agent as an active Brownian particle ~\cite{bechingerActiveParticlesComplex2016, romanczukActiveBrownianParticles2012} that moves with a constant self-propulsion speed $v_0$ along an axis $\vu{n}=(\cos{\theta},\sin{\theta})$. The position $\vb{r}$ of the particle and orientation $\theta$  of the propulsion direction evolve according to
\begin{align}
    \dv{\vb{r}}{t} &= v_0\vu{n}(\theta) + \sqrt{2D_t}\vb{\xi_t}(t) \label{eq:r} \\ 
    \dv{\theta}{t} &= \sqrt{2D_r}\xi_r(t) \label{eq:theta}
\end{align}
where $D_t$ and $D_r$ are the translational and rotational diffusion constants respectively. $\vb{\xi_t}$ and $\xi_r$ are white noise terms with unit variance. Unlike standard active Brownian motion, we consider $D_r$ to be a dynamic variable that depends on a noisy external scalar signal field $C(\vb{x}, t)$ and two internal state variables, the \textit{memory} $M$ and the \textit{volatility} $V$. The particle senses an external signal $C(\mathbf{x},t) = S(\mathbf{x}) + \xi$, where $S$ is the smooth signal field and $\xi$ is Gaussian white noise of amplitude $W$. Along its trajectory the particle compares the current signal $C(\vb{r}(t))$ to the memory $M$ to calculate a prediction error $\delta \equiv C(\vb{r}(t)) - M(t)$, then uses this error to update the memory:
\be \tau_M \dv{M}{t} = \delta(t). \label{eq:M}\ee
Here $\tau_M$ is the memory relaxation time, defining the temporal window over which the particle "remembers" past signals. At the same time, the particle adapts the volatility $V$ to the unpredictability of the environment by tracking the absolute prediction error:
\be \tau_L \dv{V}{t} = \alpha \abs{\delta(t)} - V(t) \label{eq:V}.\ee
This adaptation is controlled by the learning timescale $\tau_L$ and the adaptation sensitivity $\alpha$. The memory and volatility thus track the mean of the signal and noise amplitude respectively.

\begin{figure}
    \centering
    \includegraphics[width=\linewidth]{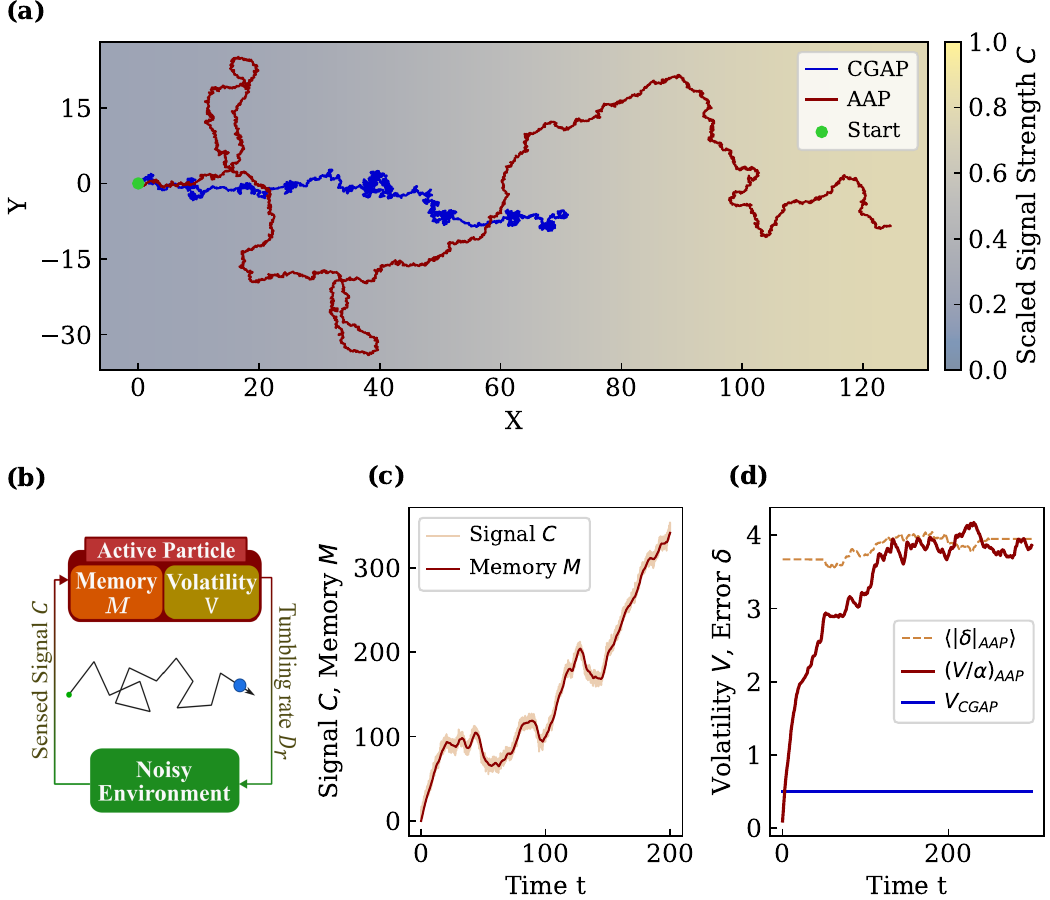}
    \caption{\textbf{(a)} Representative trajectories of the Adaptive Active Particle and the Constant-Gain Active Particle in a linear signal gradient along the $x$ direction, with additive white noise (not shown in the background coloration). The AAP travels  farther along the gradient than the CGAP in the same time. \textbf{(b)} Schematic of the model, paralleling the agent–environment loop of reinforcement learning. \textbf{(c)} Signal $C$ and Memory $M$ of the AAP over the trajectory shown in (a). $M$ tracks $C$l closely. \textbf{(d)} Volatility and prediction error for the AAP and CGAP over the trajectories shown in (a). The scaled volatility of the AAP slowly tracks the average prediction error. In contrast the volatility of the CGAP remains constant. See Supplemental Material~\cite{SupMat} for a list of parameter values.}
    \label{fig:1}
\end{figure}

The rotational diffusivity $D_r$ in Eq.~\eqref{eq:theta} is set by an internal reorientation process, which we now derive from an underlying two-state model. At each instant, the particle evaluates whether to persist in its current run or to reorient, and the rate at which it commits to reorienting defines $D_r$. This decision follows from the principle that the particle should act on its noisy sensory signal only as decisively as the reliability of that signal warrants. The particle holds one of two behavioral states, a run ($R$) or a tumble ($T$). To each state we assign a utility, which is a scalar benefit ranking outcomes, as an energy ranks configurations but with opposite sign. Persisting in a run that climbs the gradient is beneficial in proportion to the prediction error $\delta$, while reorienting carries a fixed cost $E_0$ for abandoning the current heading, $U(R) = \beta\,\delta$, $U(T) = -E_0$, with $\beta$ the coupling from prediction error to benefit. Since $\delta$ is corrupted by environmental noise, an agent that always selects the higher-utility state would reorient on every spurious fluctuation. It must instead adopt a probabilistic policy $P=\{P(R),P(T)\}$ that is selected by a single governing principle: maximize the mean utility $\langle U\rangle$ but penalize any departure from an unbiased default $P_0=\{\frac{1}{2},\frac{1}{2}\}$
that would commit the agent beyond what its information supports~\cite{ortegaThermodynamicsTheoryDecisionmaking2013}. This penalty is measured by the relative entropy $D(P\|P_0)$, giving
\begin{equation}
P^{*} = \arg\max_{P}\Big[\,\langle U\rangle - \Theta\, D(P\,\|\,P_0)\,\Big].
\label{eq:var}
\end{equation}
$\Theta$ sets the price of decisiveness. Equation~\eqref{eq:var} can be viewed as a free-energy functional, with $\langle U\rangle$ in the role of negative energy and $\Theta$ in the role of temperature. Its maximizer is the Boltzmann-like distribution over actions~\cite{SupMat}, arising here from constrained optimization rather than assumption of equilibrium. The optimal policy sets the rotational diffusivity in Eq. \eqref{eq:theta} through the odds of tumbling, $P(T)/P(R)$, giving (see Supplemental Material~\cite{SupMat}):
\begin{equation}\label{eq:D_r}
D_r = D_0\exp\!\left[-\frac{E_0+\beta\delta}{\Theta}\right],
\end{equation}
where $D_0$ is the maximum reorientation rate, reached when the utility gap $E_0+\beta\delta$ vanishes. This structure, where an agent selects actions to maximize expected reward under an information constraint, is that of a reinforcement-learning policy (Fig.~\ref{fig:1}b), here derived analytically. The scale $\Theta$ is fixed by the reliability of the agent's information, which is limited by two statistically independent sources of noise: an intrinsic floor $k_B T$ from the thermal fluctuations of the motor that executes the switch, and a sensory contribution equal to the amplitude of the environmental fluctuations that is measured through the volatility $V$~\cite{SupMat}. As independent contributions the two noise sources add, $\Theta = k_B T + V$, one term fixed and one tracking the environment.


The response gain $G$ of the policy~\cite{fairhallEfficiencyAmbiguityAdaptive2001} is the slope of the log tumbling rate with the signal, $G=\vert\partial\ln{D_r}/\partial\delta\vert=\beta/(k_B T + V)$. We call a particle whose volatility, and hence gain, is held at a constant predetermined value a \textit{Constant Gain Active Particle} (CGAP), and one that modulates its volatility according to Eq.~\eqref{eq:V} an \textit{Adaptive Active Particle} (AAP). Both the AAP and CGAP track their memory via Eq.~\eqref{eq:M} and follow the policy of Eq.~\eqref{eq:D_r} (Fig. \ref{fig:1}b). When a particle climbs the gradient, $\delta$ becomes positive and $D_r$ decays, so it reorients less often and drifts up the gradient (Fig.~\ref{fig:1}a). The memory tracks the signal along the trajectory (Fig.~\ref{fig:1}c), so this drift persists whenever the gradient is resolved. The two particles differ only in how they set the denominator $\Theta$ of Eq.~\eqref{eq:D_r}. For the CGAP $\Theta$ is fixed. For the AAP the volatility follows the running absolute prediction error scaled by the adaptation sensitivity $\alpha$ (Fig.~\ref{fig:1}d), so $\Theta$ grows with the local noise.
At high signal and noise levels ($V\gg k_BT$, $\beta\delta \gg E_0$) 
the exponent in Eq. \eqref{eq:D_r} is dominated by $\beta\delta/V$, the signal-to-noise ratio, and the AAP acts on how significant the signal is relative to the noise, rather than its magnitude. As a result the AAP travels farther up the gradient than the CGAP in a noisy field (Fig.~\ref{fig:1}a, Movie 1).

\begin{figure}
    \centering
    \includegraphics[width=1\linewidth]{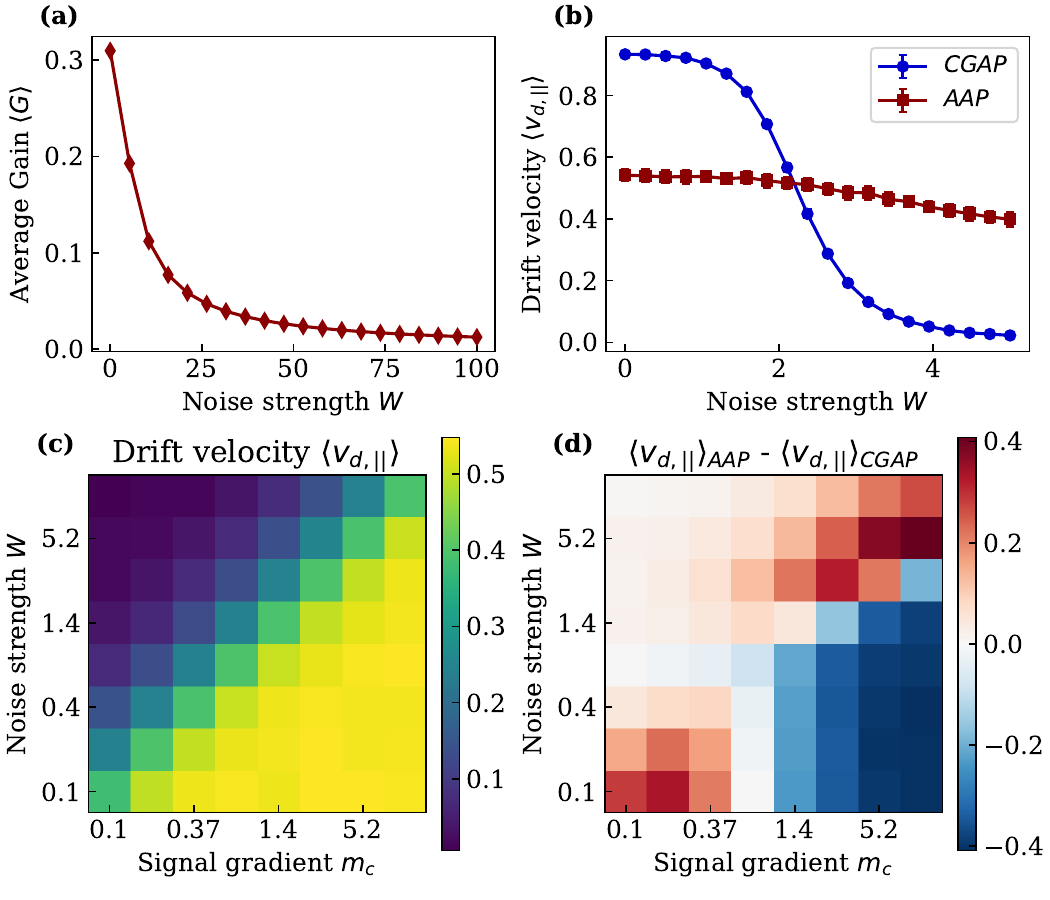}
    \caption{\textbf{(a)} The average gain $\langle G\rangle=\langle \beta/(k_BT + V)\rangle$ of the AAP decreases with the standard deviation of the Gaussian white noise in the signal. \textbf{(b)} Drift velocity along the signal gradient against noise level for the CGAP and the AAP. The CGAP performs better in quiet environments but the AAP outperforms the CGAP at high noise strengths. \textbf{(c)} Drift velocity along the signal gradient across varying noise level and signal gradient for the AAP. \textbf{(d)} Difference in drift velocity along gradient between AAP and CGAP across varying noise strength and signal gradient. For a given signal gradient, the AAP performs better than the CGAP at higher $W$.}
    \label{fig:2}
\end{figure}

\textit{Macroscopic transport in a noisy field. \textemdash} How does the adaptive policy shape chemotactic transport as environmental noise grows? To answer this we compute the mean drift velocity in a constant signal gradient $m_c$. The prediction error then splits into a deterministic bias set by the particle's orientation $\theta$ and a fluctuating environmental part, $\delta = \mu(\theta) + \eta$, $\eta \sim \mathcal{N}(0, W^2)$, where $W$ is the noise amplitude, equal to that of the sensing noise in the white-noise limit. The AAP's volatility tracks this amplitude, rising linearly with $W$, hence its response gain decreases with increasing noise (Fig.~\ref{fig:2}a). This captures the inverse scaling of response gain with stimulus variance observed in sensory systems~\cite{fairhallEfficiencyAmbiguityAdaptive2001, dahmenAdaptationStimulusStatistics2010}. On the other hand the CGAP's volatility and response gain are fixed; the two particles therefore differ only through the scale $\Theta$. Averaging the policy (Eq.~\ref{eq:D_r}) over $\eta$ shifts the baseline tumbling rate, while the orientation-dependent bias $\mu(\theta)$ sets the chemotactic response.

To gain analytical insights into the transport mechanism, we assume $\beta\mu \ll \Theta$, where the signal-induced bias is small compared to the effective temperature (our simulations make no such assumption). In this regime, the time-averaged tumbling rate is~\cite{SupMat} $\langle D_r(\theta) \rangle \approx D_\text{base}(W)\qty[1 - \chi_\text{eff}\cos\theta]$, where $D_\text{base}(W)$ is the baseline tumbling rate in the absence of a gradient and $\chi_\text{eff}$ is the effective chemotactic coupling,
\begin{align}
    D_\text{base}(W) &= D_0\exp\qty(\frac{\beta^2W^2}{2\Theta^2} - \frac{E_0}{\Theta}), \label{eq:d0}\\
    \chi_\text{eff} &= \frac{\beta v_0 m_c}{\Theta}\qty[\frac{\tau_M}{1 + D_\text{base}(W)\tau_M}]. \label{eq:chi}
\end{align}
Solving the steady-state Fokker-Planck equation for the orientation distribution~\cite{schnitzerTheoryContinuumRandom1993, SupMat} gives the drift velocity along the gradient, $v_{d,\parallel} = \tfrac{1}{2}\,v_0\,\chi_\text{eff}$. The drift grows with the gradient and falls with $D_\text{base}$, so the transport of both particles is governed entirely by how $D_\text{base}$ responds to noise.

In the noise-free limit $W \to 0$, the bias $\mu$ still fluctuates as the orientation diffuses. The AAP's volatility then tracks these self-generated signal changes, saturating at a floor $V_{\min}^{\mathrm{AAP}} \approx \tfrac{\alpha}{\sqrt{2}}\,\tau_M v_0\,\abs{m_c}$~\cite{SupMat}. 
The inability of the AAP to distinguish external noise from the signal changes its own motion produces, places a lower bound on the AAP's effective temperature $\Theta$. A CGAP tuned as a low-noise specialist, with fixed volatility $V^{\mathrm{CGAP}} < V_{\min}^{\mathrm{AAP}}$, attains a lower $\Theta$, a smaller baseline rate, and a larger coupling $\chi_\text{eff}$. The CGAP therefore outperforms the AAP in quiet environments (Fig.~\ref{fig:2}b, small $W$).

AAP outperforms the CGAP as noise grows. With its volatility rising as $V \approx \alpha\sqrt{2/\pi}\,W$, the noise term $\beta^2 W^2/2\Theta^2$ in the exponent of $D_\text{base}$ saturates and the AAP's baseline rate stays nearly constant. The CGAP has no such compensation: with $\Theta$ fixed, its baseline rate diverges as $D_\text{base} \sim e^{\beta^2 W^2/2\Theta^2}$, the kinetic damping factor $\tau_M/(1+D_\text{base}\tau_M)$ in Eq.~\eqref{eq:chi} collapses, and the drift with it. The AAP's drift instead stays substantial across a wide range of noise and gradient (Fig.~\ref{fig:2}c). The drifts of the two particles degrade in qualitatively different ways, $v_{d,\parallel}^{\mathrm{CGAP}} \sim e^{-W^2}$, and $v_{d,\parallel}^{\mathrm{AAP}} \sim 1/W$ for $W \gg \Theta/\beta$~\cite{SupMat}. The CGAP's drift collapses exponentially, while the AAP's decays only algebraically. Adaptation thus converts a catastrophic loss of chemotactic transport into a slow one, and the AAP outperforms the CGAP at every gradient once the noise is large enough (Fig.~\ref{fig:2}d).

\begin{figure}
    \centering
    \includegraphics[width=1\linewidth]{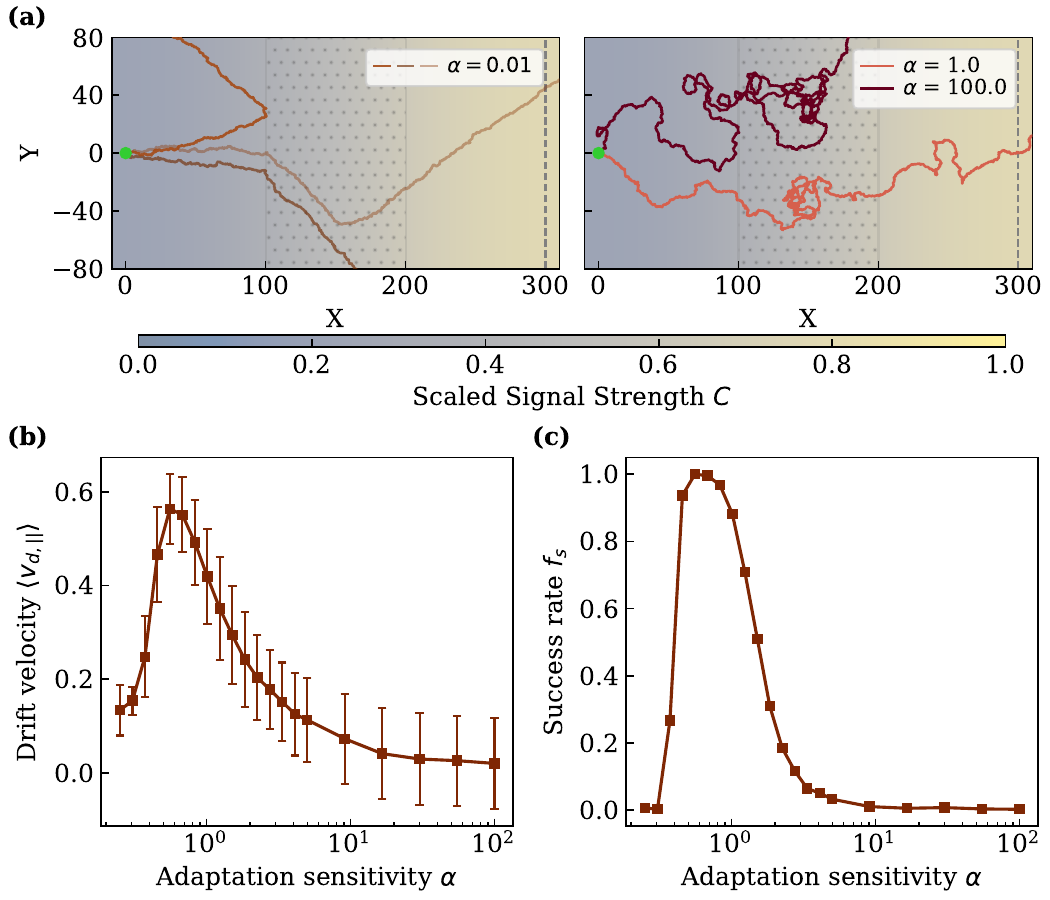}
    \caption{\textbf{(a)} Trajectories of the AAP in an alternating noise topography with varying adaptation sensitivity $\alpha$. Left panel shows multiple trajectories at low $\alpha$ and right panel shows trajectories at intermediate and high values of $\alpha$. The signal has a linear gradient in the $+x$ direction, noise amplitude $W=10.0$ in the middle textured region and $W=1.0$ everywhere else. \textbf{(b)} Drift velocity along the gradient for the AAP across varying sensitivities. $\langle v_{d,||} \rangle$ peaks around $\alpha \approx 0.6$ and decreases for very high and very low values of $\alpha$. \textbf{(c)} The success rate of particles, defined as the fraction of trajectories that cross a predefined distance from the start (dotted line in (a)) in a given time, across varying sensitivities $\alpha$.}
    \label{fig:3}
\end{figure}

\textit{Spatially varying noise. \textemdash} Noise in natural environments is not constant but fluctuates over space and time. The adaptation sensitivity $\alpha$ sets how strongly the AAP responds to the noise around it, and this controls how well it navigates through regions noisier than their surroundings. To test this, we placed the AAP in a space with alternating low- and high-noise regions (Fig.~\ref{fig:3}). As in Fig.~\ref{fig:1}, a global signal gradient points along the $x$-direction with additive white noise. In a narrow vertical strip, however, the noise is an order of magnitude higher. The particle starts to the left of this "noise barrier" and must cross it to reach the global signal maximum. 

The trajectories depend strongly on $\alpha$ (Fig.~\ref{fig:3}a, Movie 2). At intermediate $\alpha$, the particle moves more erratically inside the barrier, as expected, but is able to cross it. At very low $\alpha$, the particle responds weakly to the modest noise of the surroundings. Its volatility and rotational diffusion stay very low, producing long, straight runs (Fig.~\ref{fig:3}a, left panel). On entering the barrier, however, the much larger noise is enough to drive the volatility up even at low adaptation sensitivity, and the rotational diffusion spikes. The particle then reorients sharply in a random direction, sometimes proceeding on an extremely deviated path and occasionally appearing to be reflected back.
These abrupt turns, together with the otherwise straight runs, make the low-$\alpha$ particle inefficient in crossing the barrier (Fig.~\ref{fig:3}c). On the other hand, an AAP with a high $\alpha$
is sensitive to the noise everywhere. It tumbles frequently from the start and even more once inside the noisy strip, reorienting so often that it makes little directed progress and becomes trapped. Only at intermediate $\alpha$ the particle is responsive enough to follow the signal gradient but not so sensitive that noise traps it (Fig.~\ref{fig:3}a).

The chemotactic drift of the AAP therefore  peaks at an optimum intermediate value of $\alpha$ and is reduced at both very high and very low $\alpha$ (Fig.~\ref{fig:3}b). This is clearer in the fraction of trajectories that reach a predetermined distance along the gradient in a given time (Fig.~\ref{fig:3}c): far more particles with intermediate $\alpha$ reach the target than those with very high or very low $\alpha$.

\begin{figure}
    \centering
    \includegraphics[width=1\linewidth]{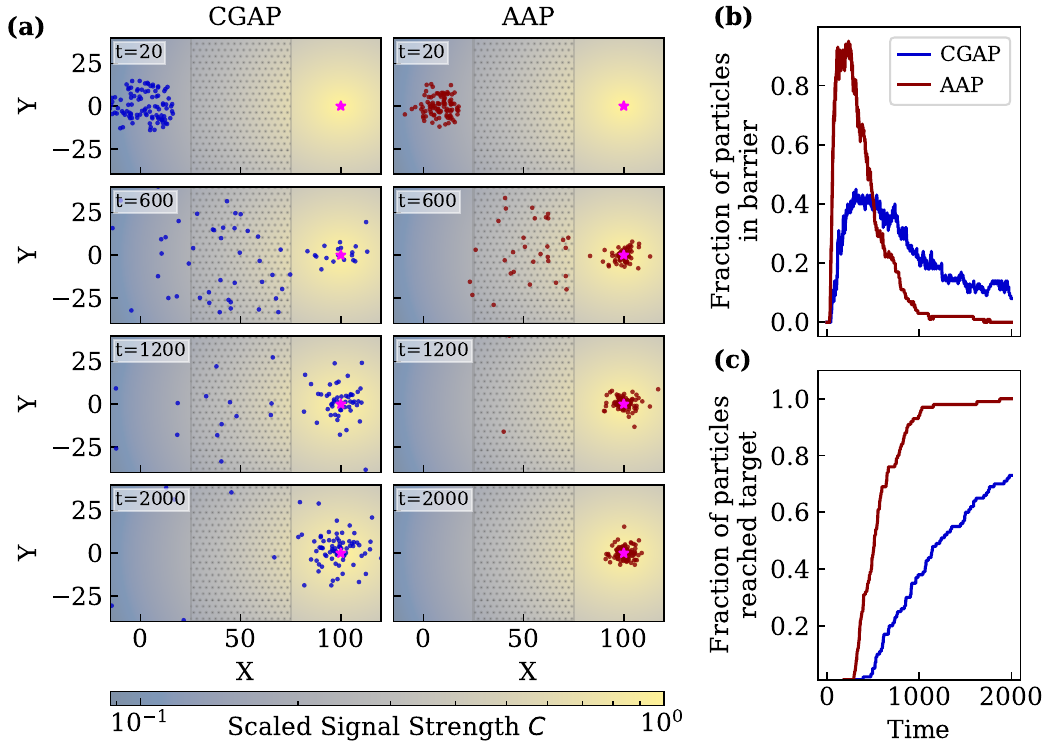}
    \caption{\textbf{(a)} Snapshots of collections of particles, CGAP (left) and AAP (right), searching for a point source of signal (magenta star) across a noise barrier (textured in gray). The signal decays exponentially from the source in the radial direction. The noise in the barrier is 20 times the surroundings. The CGAPs spend a longer time inside the barrier and form a looser cluster around the target at the end compared to the AAPs. \textbf{(b)} Fraction of particles in the barrier over time. A larger fraction of CGAPs remain caught in the barrier at higher timepoints compared to the AAPs. \textbf{(c)} Fraction of particles that reach the target over time. All AAPs reach the target in a much shorter time compared to the CGAPs.}
    \label{fig:4}
\end{figure}

\textit{Collective target finding. \textemdash} So far we have compared the transport of individual AAPs and CGAPs in linear signal gradients. We now turn to a setting closer to nature, where organisms such as bacteria move in groups and chemotactic signals emanate from localized sources. We thus simulated groups of CGAPs and AAPs searching for a point source across a noise barrier (Fig.~\ref{fig:4}, Movie 3).
The particles start as a cluster to the left of the barrier, where the noise is an order of magnitude higher than its surroundings. The signal is peaked at a point on the other side of the noise barrier and decays exponentially with distance outwards (Fig.~\ref{fig:4}a). We consider the simplest case where the particles interact solely through steric repulsion. We find that collections of both CGAPs and AAPs cross the barrier and accumulate around the signal peak (Fig. \ref{fig:4}a). However, the CGAPs spend a longer time in the barrier (Fig \ref{fig:4}b) and take longer to reach the target (Fig. \ref{fig:4}c) compared to the collection of AAPs.

\textit{Discussion. \textemdash} We introduced the Adaptive Active Particle, a minimal model in which rotational diffusivity is set by an internally estimated signal-to-noise ratio. This variance-adaptive policy confers large performance gains in noisy environments at modest cost in quiet ones, provided the adaptation sensitivity is tuned to an optimum value. The advantage persists at the population level, where AAPs cross localized noise barriers to a target faster than fixed-gain particles.
 
Our results refine recent perspectives on noise in bacterial chemotaxis. Endres~\cite{endres2026robust} showed that fixed-gain run-and-tumble chemotaxis is robust to small noise, consistent with our model in that regime. At high noise, beyond the scope of that analysis, we find instead that the drift of a fixed-gain agent collapses, and that variance adaptation is required to sustain it. Our approach is likewise deterministic in contrast to the stochastic framework of Karin and Alon~\cite{karinTemporalFluctuationsChemotaxis2021}. Rather than randomly switching pathway gain between exploratory and exploitative regimes, our agent continuously tracks environmental noise and adjusts its effective temperature along its trajectory.

A further consequence of variance adaptation is that our policy naturally reproduces Weber's law~\cite{FechnerElemente1860, pardo-vazquezMechanisticFoundationWebers2019} in the limit when the signal-driven terms dominate the fixed energy costs. Because the memory $M$ and volatility $V$ are both linear filters of the sensed signal, a rescaling of the signal $C \to \kappa C$ rescales the prediction error and the volatility in equal proportion, $\delta \to \kappa\delta$ and $V \to \kappa V$. The policy exponent $\beta\delta/(k_BT + V)$ is then invariant whenever $E_0\ll \beta \delta$ and $k_BT\ll V$, so the particle responds to relative rather than absolute changes in the signal. This fold-change detection~\cite{goentoroEvidenceThatFoldChange2009} is a hallmark of bacterial chemotaxis and its logarithmic sensing~\cite{kalininLogarithmicSensingEscherichia2009}.
 
Two extensions would bring the model closer to real chemotactic agents. First, we treat adaptation as costless, whereas real adaptive processes are dissipative and face an energy-speed-accuracy trade-off~\cite{lanEnergySpeedAccuracy2012}. Second, we hold the signal field fixed, whereas real agents deplete the chemoattractants they follow, coupling their trajectory to the signal landscape~\cite{weijerChemotaxisActiveDegradation2020}.

\end{document}